\documentclass{main}
\usepackage{mathrsfs}
\usepackage{amssymb,bm}
\usepackage{amsmath}
\usepackage{mathtools}
\usepackage{slashed}
\usepackage{physics}
\usepackage{hyperref}

\def\be{\begin{equation}}
\def\ee{\end{equation}}
\def\bea{\begin{eqnarray}}
\def\eea{\end{eqnarray}}

\newcommand{\Pt}{\mathbf{P}_{\perp}}

\newcommand{\aem}{\alpha_\mathrm{em}}

\newcommand{\jpsi}{$\mathrm{J}/\psi$ }
\newcommand{\jpsim}{\mathrm{J}/\psi}
\newcommand{\bt}{\mathbf{b}}
\newcommand{\rt}{\mathbf{r}}
\newcommand{\Bt}{\mathbf{B}}
\newcommand{\xpom}{x_\mathbb{P}}

\newcommand{\xt}{\mathbf{x}}
\newcommand{\qt}{\mathbf{q}}
\newcommand{\kt}{\mathbf{k}}

\newcommand{\Mcal}{\mathcal{M}}
\newcommand{\Fcal}{\mathcal{F}}
\newcommand{\Acal}{\mathcal{A}}
\newcommand{\der}{\mathrm{d}}

\newcommand{\Photo}{}

\begin{document}

\title{\Large Vector meson production in ultraperipheral heavy ion collisions}

\author{Bj\"orn Schenke \footnote{speaker, bschenke@bnl.gov}}
\address{Physics Department, Brookhaven National Laboratory, Upton, NY 11973, USA}

\author{Heikki M\"antysaari}
 \address{Department of Physics, University of Jyv\"askyl\"a, P.O. Box 35, 40014 University of Jyv\"askyl\"a, Finland\\ Helsinki Institute of Physics, P.O. Box 64, 00014 University of Helsinki, Finland}

\author{Farid Salazar}
\address{Institute for Nuclear Theory, University of Washington, Seattle WA 98195-1550, USA}

\author{Chun Shen}
 \address{Department of Physics and Astronomy, Wayne State University, Detroit, Michigan 48201, USA\\ RIKEN BNL Research Center, Brookhaven National Laboratory, Upton, NY 11973, USA}

\author{Wenbin Zhao}
 \address{Nuclear Science Division, Lawrence Berkeley National Laboratory, Berkeley, California 94720, USA\\ Physics Department, University of California, Berkeley, California 94720, USA}

\maketitle

\abstracts{We review model calculations of exclusive vector meson production in ultraperipheral heavy ion collisions. We highlight differences and similarities between different dipole models and leading twist shadowing calculations. Recent color glass condensate calculations are presented with focus on effects from nuclear structure and azimuthal anisotropies driven by interference effects.}


\section{Introduction}
Ultraperipheral heavy ion collisions (UPCs) at the Relativistic Heavy Ion Collider (RHIC) and the Large Hadron Collider (LHC) provide access to photonuclear events at high energy. They allow the study of processes that are otherwise only accessible by an electron- (or muon-) ion collider. 
While the kinematics is less controlled and versatile, because we cannot measure the momentum of a scattered electron, and the $Q^2$ is constrained to values close to zero, UPCs have the advantage of being able to probe very small momentum fractions $x$ in the target. This is particularly useful for accessing the gluon saturation regime and exploring features of non-linear Quantum Chromodynamics (QCD) in dense systems \cite{Morreale:2021pnn}.

One interesting process that is sensitive to the spatial parton distributions in the target as well as to saturation effects in the gluon distribution is the exclusive production of vector mesons. For coherent diffractive vector meson production, in which the target remains intact, measurements differential in the transverse momentum transfer squared $|t|$ contain information on the transverse (to the beam line) spatial structure of the target. In the case of incoherent production, for which the target breaks up, one is sensitive to fluctuations of its geometry \cite{Miettinen:1978jb,Mantysaari:2020axf}. 

Both in e+A scattering and UPCs, part of the process can be understood as a virtual photon interacting with the target. In UPCs, the photon is radiated from the moving projectile nucleus. Consequently, it is almost real, as $Q^2\sim 1/R_A^2$ with $R_A$ the nuclear radius. The fact that both nuclei can be either the photon source or the target can complicate things and interference between the two scenarios must be taken into account. 

We will focus in detail on the coherent production of $J/\psi$ vector mesons in UPCs of Pb nuclei at the LHC and discuss a variety of models and how nuclear effects are taken into account. We will point out some similarities and differences between the models, and present comparisons to experimental data. We will further discuss calculations within the Color Glass Condensate (CGC) framework in some more detail and highlight results for the energy dependence of nuclear suppression, azimuthal anisotropies caused by interference effects, and the effects of nuclear structure on diffractive vector meson production. 

\section{General considerations}
The photon flux is given by  \cite{Bertulani:2005ru} 
\begin{equation}
\label{eq:bint_flux}
N(\omega_\pm) = \int_{|\Bt|>B_\text{min}} \!\!\!\!\!\!\! \dd[2]{\Bt} n(\omega_\pm, \Bt) \,,
\end{equation}
where $\Bt$ is the impact parameter vector between the centers of the two nuclei, 
$\omega_\pm = (M_V/2) e^{\pm y}$, with $M_V$ the vector meson mass, and
\begin{equation}
\label{eq:flux}
    n(\omega,\Bt) = \frac{Z^2 \aem \omega^2}{\pi^2 \gamma^2}  K_1^2\left( \frac{\omega |\Bt|}{\gamma}\right).
\end{equation}
Here, $\aem = 1/137$ the fine-structure constant, $Z$ is the ion charge and $\gamma=A\sqrt{s}/(2M_A)$, with $M_A$ the mass of the nucleus. $B_{\rm min}$ is the minimal impact parameter to not have a hadronic interaction. It has to be on the order of $2\,R_A$, with $R_A$ the nuclear radius.

At midrapidity, the cross section for the process $A_1+A_2\rightarrow V+A_1+A_2$ is given by
\begin{equation}
    \left.\frac{\dd\sigma^{A_1+A_2\rightarrow V+A_1+A_2}}{\dd|t|dy}\right|_{y=0} = 2\int \dd[2]\mathbf{B} \, n(\omega,|\mathbf{B}|) \frac{\dd\sigma^{\gamma+A\rightarrow V+A}}{\dd|t|} [1-\cos(\boldsymbol{\Delta}\cdot \mathbf{B})]\,\theta(|\mathbf{B}|-2R_A)\,,
\end{equation}
where $\boldsymbol{\Delta}^2 = -t$. This shows that at midrapidity interference leads to a vanishing differential cross section at $t=0$.

This simplified expression neglects photon $\kt$, which is a good approximation except around diffractive minima in the coherent cross section, because $\kt^2 \lesssim Q^2 \sim 1/R_A^2$. In coordinate space it is realized by assuming $|\Bt| \gg |\bt|$, with $\bt$ the impact parameter vector of the quasi-real photon relative to the center of the target nucleus.
Further neglecting interference one obtains for the rapidity-dependent result
\begin{equation}
\label{eq:dsigma_dy_nointerf_nokt}
   \frac{\der \sigma^{A_1+A_2 \to V+A_1+A_2}}{\dd{y}} =  N(\omega_{+}) \sigma_{+}^{\gamma^* + A \to V + A}
   +  N(\omega_{-}) \sigma_{-}^{\gamma^* + A \to V + A}\,.
\end{equation}

Here, $\sigma_{+}^{\gamma^* + A \to V + A}$ and $\sigma_{-}^{\gamma^* + A \to V + A}$ refer to the photon-nucleus cross sections where the target structure is probed at different longitudinal momentum fractions $\xpom = (M_V/\sqrt{s})e^{\mp y}$.

\begin{figure}
    \centering
    \includegraphics[scale=0.5]{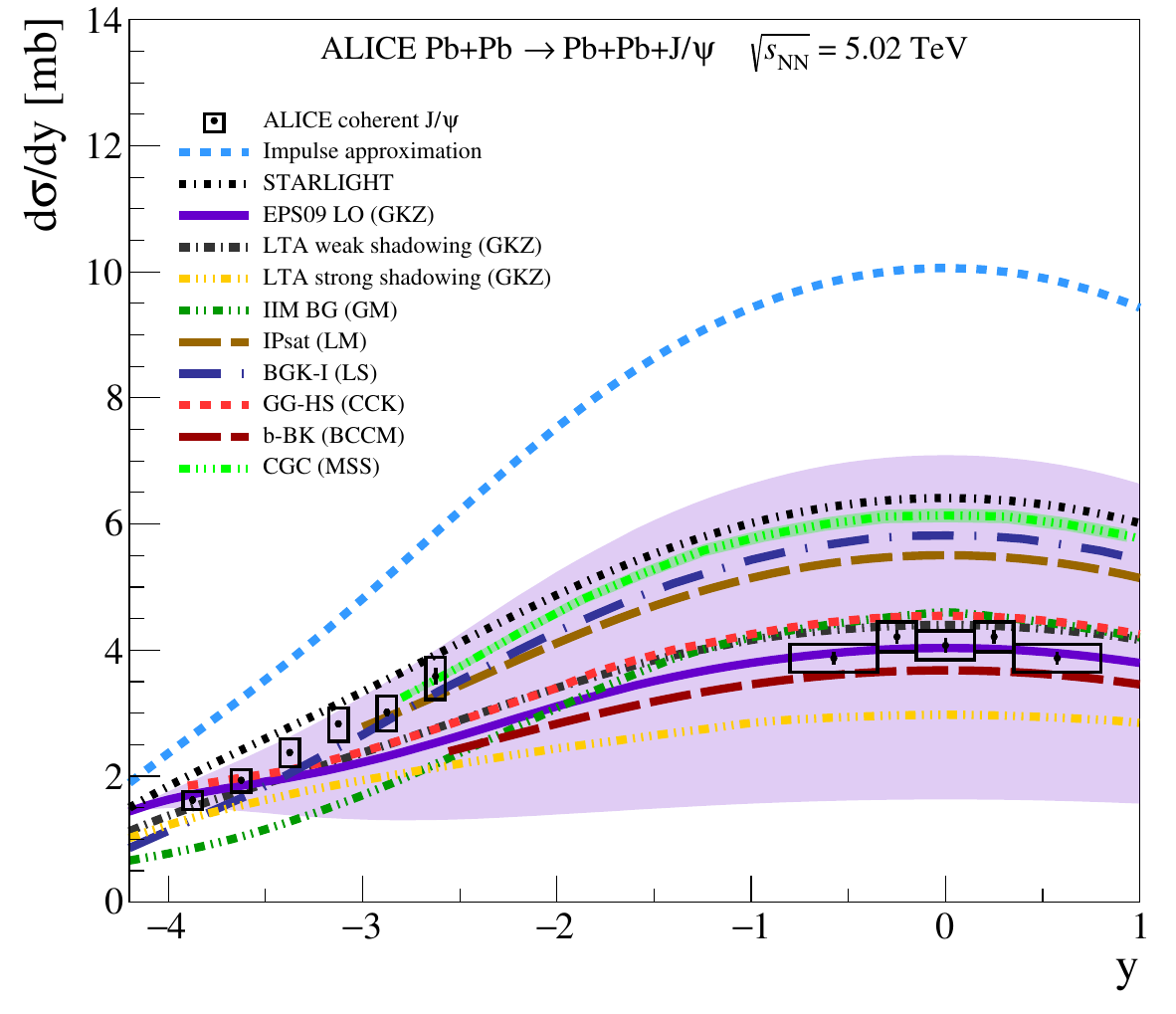}
    \caption{Differential cross section of coherent diffractive $J/\psi$ photoproduction in Pb+Pb UPC events. For the ALICE data \protect\cite{ALICE:2019tqa,ALICE:2021gpt}, the error bars (boxes) show the statistical (systematic) uncertainties. The theoretical calculations are discussed in the text. The purple band represents the uncertainties of the EPS09 LO calculation. \label{fig:ALICEplot}}
\end{figure}

\section{Model comparisons}
The difference between the theoretical models shown in Fig.\,\ref{fig:ALICEplot} lies mainly in how the cross section $\sigma_{\pm}^{\gamma^* + A \to V + A}$ is determined. We separate the models into two classes, the leading order (LO) pQCD description following Ryskin \cite{Ryskin:1992ui} and the dipole picture, first proposed in \cite{Kopeliovich:1981pz}.

\subsection{LO pQCD framework}
In the LO pQCD framework, the cross section takes on the form \cite{Guzey:2013qza}
\begin{align}\label{eq:lopqcd}
    \sigma^{\gamma^* + A \to V + A} 
    &=\frac{C_A(\mu^2)}{C_p(\mu^2)}\frac{d\sigma^{\gamma^{*} + p \to V + p}(W_{\gamma p}, t=0)}{dt} \left[\frac{xg_A(x,\mu^2)}{A x g_p(x,\mu^2)}\right]^2\Phi_A(t_{\rm min})\,,
\end{align}
where $x g_p$ and $x g_A$ are the gluon distributions in the proton and nucleus, respectively, and $\mu$ is a hard scale on the order of the charm quark mass (for $J/\psi$ production). 
The cross section at $t=0$ for a proton target is given by
\begin{equation}
    \frac{\dd\sigma^{\gamma^*+p\rightarrow V+p}(W_{\gamma p}, t=0)}{\dd t} = \frac{1}{16 \pi} |A^{\gamma^*+p\rightarrow V+p}|^2\,,
\end{equation}
where
\begin{equation}\label{eq:LTAAmp}
    A^{\gamma^*+p\rightarrow V+p} = i 4\pi^2 \sqrt{\frac{M_V^3 \Gamma_{ee}}{48 \alpha \mu^8}} \alpha_s(\mu^2) xg_p(x,\mu^2) F(\mu^2) \sqrt{1+\eta^2}R_g\,.
\end{equation}

The function $\Phi_A(t_{\rm min}) = \int_{-\infty}^{t_{\rm min}} \dd t |F_A(t)|^2$, with the nuclear form factor $F_A$ and $t_{\rm min}=-M_V^4 m_N^2/W_{\gamma p}^4$ is the minimal squared momentum transfer to the nucleus. The function $F(\mu^2)$, that lies between 0 and 1, contains effects that go beyond the leading order collinear factorization used, for example next-to-leading order corrections, corrections associated with the charmonium wave function, and power-suppressed corrections in the overlap of the photon and $J/\psi$ wave functions. \cite{Guzey:2013qza} The factor $C_A/C_p = (1+\eta_A^2)\bar{R}_{g,A}^2/[(1+\eta^2)\bar{R}^2_{g}]$. Here, $\eta_A$ and $\eta$ are the ratios of the real to imaginary part of the $\gamma+A\rightarrow V+A$ and $\gamma+p\rightarrow V+p$ scattering amplitude, respectively, and $R_{g,A}$ and $R_g$ are ``skewness factors'' that correct for the fact that we are not using generalized parton distributions (GPDs) but the usual ones. 
One determines $\eta_A$ and $\bar{R}_{g,A}$ using the asymptotic small-$x$ dependence of the nuclear gluon distribution. \cite{Guzey:2013qza}
This factor introduces some nuclear modification, as $R_{g,A}$ differs from $R_{g}$, but the main shadowing effect is governed by the ratio of gluon distributions
\begin{equation}
    R=\frac{xg_A(x,\mu^2)}{A x g_p(x,\mu^2)}\,.
\end{equation}

The ``impulse approximation'' in Fig.\,\ref{fig:ALICEplot} is obtained using Eq.\,\eqref{eq:lopqcd} with $R=1$ and $C_A/C_p=1$. \cite{Guzey:2013qza} It clearly overshoots the data at all rapidities. In the leading twist approximation (LTA), $R$ is obtained from the nuclear gluon distribution in leading twist shadowing (and so is $C_A/C_p\approx 0.9$) \cite{Frankfurt:2003qy}
\begin{align}\label{eq:gALTA}
    g_A(x,\mu^2)=Ag_p(x,\mu^2)-8\pi \Re\Bigg[ & \frac{(1-i\eta)^2}{1+\eta^2} \int \dd[2] \bt  \int_{-\infty}^{\infty} \dd z_1 \int_{-\infty}^{\infty} \dd z_2 \int_x^{x_{\mathbb{P}}^0} \dd x_{\mathbb{P}} \, g_N^D(x/x_{\mathbb{P}},x_{\mathbb{P}},\mu^2,t_{\rm min})\, \notag\\ &\rho_A(\boldsymbol{b},z_1) 
    \rho_A(\boldsymbol{b},z_2) e^{ix_{\mathbb{P}} m_N (z_1-z_2)} e^{-\frac{1}{2} \sigma_{\rm eff}(x,\mu^2) (1-i\eta_A)\int_{z_1}^{z_2} \dd z \rho_A(\boldsymbol{b},z)}\Bigg]\,,
\end{align}
where $g_N^D(x/x_{\mathbb{P}},x_{\mathbb{P}},\mu^2,t_{\rm min})$ is the diffractive gluon density of the nucleon, $x_{\mathbb{P}}$ is the pomeron momentum fraction, and $x_{\mathbb{P}}^0=0.03$ a cutoff parameter. This includes the interaction with two nucleons at longitudinal positions $z_1$ and $z_2$, and the interaction with three or more nucleons is absorbed into the attenuation factor, the exponential including $\sigma_{\rm eff}$, which is the effective cross section for the elastic rescattering of the produced diffractive state. The uncertainty in $\sigma_{\rm eff}$ leads to the variation between `LTA weak shadowing' and `LTA string shadowing' in Fig.\,\ref{fig:ALICEplot}.
The factor $e^{ix_{\mathbb{P}} m_N (z_1-z_2)}$ takes into account longitudinal momentum transfer, or a finite coherence length. In the EPS09 LO curve, the factor $R$ is obtained from the nuclear PDFs obtained using global fits of available data on lepton-nucleus DIS and hard scattering with nuclei at the Tevatron and LHC. \cite{Eskola:2009uj}

The STARLIGHT result shown in Fig.\,\ref{fig:ALICEplot} includes Glauber-like rescattering. Here, we have similar to the impulse approximation
\begin{equation}\label{eq:sigmaIA}
    \sigma^{\gamma^* + A \to V + A} = \left.\frac{\dd \sigma^{\gamma^* + A \to V + A}}{\dd |t|}\right|_{t=0} \Phi_A(t_{\rm min}) \,.
\end{equation}
Using the optical theorem and vector meson dominance \cite{Bauer:1977iq} one can write
\begin{equation}
    \left.\frac{\dd \sigma^{\gamma^* + A \to V + A}}{\dd|t|}\right|_{t=0}  = 
    \frac{4 \pi \alpha}{f_v^2}\left.\frac{\dd \sigma^{V + A \to V + A}}{\dd |t|}\right|_{t=0} = \frac{\alpha \sigma_{\rm tot}^2(VA)}{4f_v^2}\,,
\end{equation}
where $\alpha$ is the electromagnetic coupling constant, $e^2/\hbar c$, and $f_v$ is the vector meson-photon coupling. \cite{Klein:1999qj}
The total $V+A\rightarrow V+A$ cross section follows from a Glauber calculation
\begin{align}\label{eq:stot}
    \sigma_{\rm tot}(VA) = 2 \int \dd[2]{\bt} \left(1-e^{\sigma_{\rm tot}(Vp)T_{A}(\boldsymbol{b})/2}\right)\,,
\end{align}
where $T_A(\boldsymbol{b}) = \int \dd z \rho_A(\boldsymbol{b},z)$ is the nuclear thickness function.\footnote{In practice, STARLIGHT used the inelastic cross section $ \sigma_{\rm inel}(VA) = \int \dd[2]{\bt} \left(1-e^{\sigma_{\rm tot}(Vp)T_{A}(\boldsymbol{b})}\right)$ instead of \eqref{eq:stot}.\cite{Frankfurt:2015cwa}} Throughout this text we assume $\int \dd[2]{\bt}\, T_A(\boldsymbol{b}) = A$, which may differ from some of the references below.
Just like for nuclei above, we have
\begin{equation}
        \sigma_{\rm tot}^2(Vp) = \frac{4 f_v^2}{\alpha} \left.\frac{\dd\sigma^{\gamma^*+p\rightarrow V+p}}{\dd |t|} \right|_{t=0}\,,
\end{equation}
and $\dd \sigma^{\gamma+p\rightarrow V+p}/\dd t|_{t=0}$ is determined from experimental data. \cite{Klein:1999qj}

\subsection{Dipole models}
Next, we move on to the various dipole models shown in Fig.\,\ref{fig:ALICEplot}. Here, the picture is that we are at high energy and the virtual photon first splits into a quark anti-quark dipole, which subsequently interacts with the target.
A measure of whether the dipole picture is appropriate is the coherence time $l_c = 2\nu/(Q^2+M^2_{q\bar{q}})$, compared to the nuclear size scale ($\nu$ is photon energy in the target rest frame, and $M^2_{q\bar{q}} = (m_q^2+k_T^2)/z(1-z)$). In the limit of very small $x=Q^2/2m_N\nu$ the coherence length is longer than the nuclear size and the ``frozen'' dipole picture is appropriate. If the coherence time is smaller than the nuclear size scale one should correct for dipole size fluctuations during the propagation in the nucleus, which corresponds to inclusion of the phase shifts between DIS amplitudes on different nucleons, as they are included in Eq.\,\eqref{eq:gALTA}. \cite{Kopeliovich:2016jjx} For $x\leq 10^{-2}$, the factor $e^{i(z_1-z_2)m_N x_\mathbb{P}}$ can be safely set to unity. \cite{Frankfurt:2011cs}

For the coherent process that we are discussing here, the cross section is given by \cite{Kopeliovich:1991pu}
\begin{equation}\label{eq:cohcross}
\frac{\dd\sigma^{\gamma^*+A\rightarrow V+A}}{\dd |t|} = \frac{1}{16 \pi} | A(x,Q^2,\boldsymbol{\Delta})|^2\,,
\end{equation}
where the dipole amplitude is
\begin{equation}\label{eq:Amp}
    A(x,Q^2,\boldsymbol{\Delta}) = i \int \dd[2]\rt \int \dd[2]\bt \int \frac{\dd z}{4\pi} (\psi^*\psi_V)(Q^2,\boldsymbol{r},z) e^{-i(\boldsymbol{b}-(1/2-z)\boldsymbol{r})\cdot \boldsymbol{\Delta}} \frac{\dd\sigma_{\rm dip}}{\dd[2]\bt} (\bt, \rt, x)\,.
\end{equation}
Here, $(\psi^*\psi_V)(Q^2,\boldsymbol{r},z)$ represents the wave function overlap of the photon and vector meson wave functions, and $\frac{\dd\sigma_{\rm dip}}{\dd[2]\bt} (\bt, \rt, x)$ is the average dipole cross section (an average over configurations is performed for models that include explicit fluctuations).

Considering the proton target, we note that the dipole model amplitude \eqref{eq:Amp} can be brought into the form Eq.\,\eqref{eq:LTAAmp} without the real part and skewness corrections, by taking the hard scattering (small $r$) and non-relativistic (for which $z=0.5$) limits. \cite{Kopeliovich:2016jjx,Anand:2018zle} In that case, one should take
\begin{equation}
    \frac{\dd\sigma_{\rm dip}}{\dd[2]\bt} = \frac{\pi^2}{N_c} r^2 \alpha_s(\mu^2) x g (x,\mu^2) T(\bt)\,,
\end{equation}
essentially the small $r$ limit of the BGK or IPSat model, which are discussed for nuclei below.

For the first dipole model shown in Fig.\,\ref{fig:ALICEplot}, IIM BG (GM) \cite{SampaiodosSantos:2014puz}, the dipole cross section for a nuclear target is 
\begin{equation}\label{eq:dsdipdb}
    \frac{\dd\sigma_{\rm dip}}{\dd[2]\bt} = 2(1-e^{-\frac{1}{2} \sigma^{\rm IIM\,BG}_{\rm dip}(x, \rt) T_A(\bt)})\,,
\end{equation}
and $\sigma^{\rm IIM\,BG}$ is a parametrization fit to HERA data from Iancu, Itakura, and Munier (IIM) \cite{Iancu:2003ge,Rezaeian:2013tka}
\begin{equation}\label{eq:IIMBG}
    \sigma_{\rm dip}^{\rm IIM\,BG}(x,r) = \sigma_0
\begin{cases}
0.7\left(\frac{\bar{\tau}^2}{4}\right)^{\gamma_{\rm eff} (x, r)}, \text{for } \bar{\tau}\leq 2, \\
1-\exp[-a \ln^2(b\bar{\tau})], \text{for } \bar{\tau}>2\,,
\end{cases}
\end{equation}
where $\bar{\tau} = r Q_s(x)$, with the saturation scale $Q_s=(x_0/x)^{\lambda/2}$, and $\sigma_0=2\pi R_p^2$, with $R_p$ the proton radius. Here, $\gamma_{\rm eff}(x,r) = \gamma_{\rm sat}+ \ln(2/\bar{\tau})/(\kappa \lambda y)$ with $\kappa = 9.9$. For the vector meson wave function a boosted Gaussian (BG) is used. The structure of Eq.\,\eqref{eq:dsdipdb} for a nuclear target follows from the Glauber-Gribov methodology proposed in \cite{Armesto:2002ny} and is common to almost all dipole models discussed here, except the b-BK model. 

In the next dipole model, BGK-I (LS), Eq.\,\eqref{eq:dsdipdb} is replaced with an expression that takes the real part of the dipole-nucleon amplitude into account and uses for the dipole cross section \cite{Luszczak:2019vdc}
\begin{equation}
 \sigma_{\rm dip}^{\rm BGK-I}(x,r) = \tilde{\sigma}_0 \left(1-\exp\left[-\frac{\pi^2r^2\alpha_s(\mu^2)xg(x,\mu^2)}{3\tilde{\sigma}_0}\right]\right)\,,
\end{equation}
where $xg(x,\mu^2)$ is obtained from DGLAP evolution of the initial condition $xg(x,\mu_0^2) = A_g x^{-\lambda_g} (1-x)^{C_g}$. Here, the free parameters are $\tilde{\sigma}_0$, $\mu_0^2$, $A_g$, $\lambda_g$, and $C_g$.

In the IPSat (LM) model \cite{Lappi:2013am} calculation shown, the authors assumed a large and smooth nucleus, leading to an expression that corresponds to replacing $\sigma^{\rm IIM\,BG}_{\rm dip}(x, \boldsymbol{r})$ in $\eqref{eq:dsdipdb}$ by
\begin{equation}\label{eq:sigmaIPSat}
    \sigma_{\rm dip}^{\rm IPSat} = 4 \pi B_p [1-\exp(-r^2 F(x,r))]\,,
\end{equation}
with
\begin{equation}
    F(x,r) = \frac{1}{2\pi B_p} \frac{\pi^2}{2N_c} \alpha_s\left(\mu_0^2+\frac{C}{r^2}\right) xg\left(x,\mu_0^2+\frac{C}{r^2}\right)\,.
\end{equation}
Here, $C$ is chosen to be 4 and $\mu_0^2=1.17\,{\rm GeV}^2$ is the result of a fit to HERA data. \cite{Kowalski:2006hc} The $r$ dependence of the gluon distribution is obtained from DGLAP evolution of the initial condition given by the same $   xg(x, \mu_0^2)$ as in the BGK-I model above.

The next dipole model, GG-HS \cite{Cepila:2017nef}, has a similar structure as IPSat above, but is based on a Golec-Biernat - W\"usthoff (GBW) dipole amplitude. \cite{Golec-Biernat:1998zce}

So, once again we start from Eq.\,\eqref{eq:dsdipdb} (hence the GG = Glauber-Gribov in the name of this model) but this time replace $\sigma^{\rm IIM\,BG}_{\rm dip}(x, \boldsymbol{r})$ by 
\begin{equation}
     \sigma_{\rm dip}(x,r) =  \sigma_0 [1 - \exp(-r^2 Q_s^2(x)/4)]\,,
\end{equation}
where $Q_s^2 = Q_{s0}^2 (x_0/x)^\lambda$. This model implements subnucleonic hot spots (HS), $N_{\rm hs}$ of them per nucleon, each with a Gaussian density distribution of width $B_{\rm hs}$.
Hot spots have no direct effect on the coherent cross section as that is sensitive to the average structure of the target. The incoherent cross section, in particular at large $|t|$ is sensitive to hot spots. \cite{Cepila:2017nef,Mantysaari:2017dwh}

Finally, in the b-BK (BCCM) model \cite{Bendova:2020hbb} uses expression \eqref{eq:Amp} with 
\begin{equation}
    \frac{\dd\sigma_{\rm dip}}{\dd[2]\bt} = 2 N(\rt,\bt,x)\,,
\end{equation}
where $N$ is evolved with impact parameter dependent Balitsky-Kovchegov (BK) evolution \cite{Balitsky:1995ub,Kovchegov:1999yj}, starting from an initial condition for the nucleus with the impact parameter dependence given by the Woods-Saxon distribution for Pb. \cite{Cepila:2020xol} An alternative approach that employs the BK evolved dipole cross section for the proton embedded in a Glauber-Gribov expression as used above was also studied \cite{Bendova:2020hbb}. It leads to less suppression than the evolution of the nuclear dipole cross section, as the latter leads to stronger saturation effects.

It is noteworthy that none of the models achieves a good simultaneous description of the experimental data at forward and mid-rapidities.

\subsection{Color Glass Condensate}
The Color Glass Condensate (CGC) calculation is also based on the dipole picture, so we employ expression \eqref{eq:Amp} with
\begin{equation}\label{eq:CGCdip}
    \frac{\dd\sigma_{\rm dip}}{\dd[2]\bt} = \langle N(\boldsymbol{r},\boldsymbol{b},x) \rangle_\Omega = \langle 2 N(\boldsymbol{x}-\boldsymbol{y},(\boldsymbol{x}+\boldsymbol{y})/2,x) \rangle_\Omega = 1 - \langle{\rm Tr} (V(\boldsymbol{x})V^\dag(\boldsymbol{y}))/N_c\rangle_\Omega\,,
\end{equation}
with the light-like Wilson line 
\begin{equation}
  V(\xt) = \mathrm{P}_{-}\left\{ \exp\left({-ig\int_{-\infty}^\infty \dd{x^{-}} \frac{\rho^a(x^-,\xt) t^a}{\boldsymbol{\nabla}^2 - m^2} }\right) \right\}\,.
  \label{eq:wline_regulated}
\end{equation}
$\mathrm{P}_{-}$ represents path ordering in the $x^-$ direction, and $m$ regulates unphysical Coulomb tails. 
Color charges $\rho^a$ (with color index $a$) are sampled from a distribution whose width is given by the average squared color charge density, which can be obtained from its relation to the local saturation scale extracted from the IPSat amplitude above \cite{Schenke:2012wb,Schenke:2012hg}.
This corresponds to employing a modified McLerran-Venugopalan (MV) model \cite{McLerran:1993ni,McLerran:1993ka} that includes the geometry of the target. Note that we included an explicit average over configurations $\Omega$.

As can be seen from Eq.\,\eqref{eq:dsdipdb} (along with \eqref{eq:sigmaIPSat} for $\sigma_{\rm dip}$), in the IPsat model the local saturation scale $Q_s^2(\xt)$ is proportional to the local transverse density $T_{p}(\xt)$.  For nuclei, one first samples nucleon positions from a Woods-Saxon distribution, and then calculates the total density by summing the nucleon density profiles \cite{Schenke:2012wb}. When constraining the geometry at an initial $x_0$, the evolution to smaller $x$ via the JIMWLK equations \cite{Mueller:2001uk} should be included \cite{Mantysaari:2023xcu}. This is done for the CGC (MSS) curve shown in Fig.\,\ref{fig:ALICEplot}, which also includes subnucleon fluctuations. The details of the implementation are described in the corresponding publication \cite{Mantysaari:2022sux}.

Incoherent diffraction, which we have not discussed in detail so far, is the diffractive process in which the target breaks up. Its cross section has the form of a variance of the scattering amplitude and is thus sensitive to fluctuations. \cite{Miettinen:1978jb}
As the CGC framework includes fluctuations of color charges and nuclear geometry, including nucleon and subnucleon fluctuations, it is well suited to address incoherent diffraction. The differential cross section for incoherent diffractive vector meson production is given by
\begin{equation}
    \frac{\dd \sigma^{\gamma^* + A \rightarrow V+A^*}}{\dd t}=\frac{1}{16\pi}\left(\left\langle\left| A\left(x,Q^2,\boldsymbol{\Delta}\right)\right|^2\right\rangle_\Omega-\left|\left\langle A\left(x,Q^2,\boldsymbol{\Delta}\right)\right\rangle_\Omega\right|^2\right)\,,
\end{equation}
with $A$ from Eq.\,\eqref{eq:Amp} but with the modification that we do not include the averages $\langle\cdot\rangle_\Omega$ in $\eqref{eq:CGCdip}$. Instead, we have included the averages over configurations in this expression, to allow for the computation of the variance of $A$.

\section{Nuclear suppression}
To quantify the magnitude of saturation effects in $J/\psi$ photoproduction, one can compute nuclear suppression factors separately for the coherent and incoherent channels. We define the suppression factor for the coherent production as
\begin{equation}
    S_\mathrm{coh} = \frac{\sigma^{\gamma+A\rightarrow V+A}}{\sigma^\mathrm{IA}},
\end{equation}
where $\sigma^\mathrm{IA}$ is given by Eq.\,\eqref{eq:sigmaIA}, the $\gamma+p$ result scaled to the $\gamma+\mathrm{Pb}$ case by only taking into account the nuclear form factor $F(t)$ and neglecting all other potential nuclear effects. For LHC kinematics, one can set $t_\mathrm{min}=0$. Note that in some studies $S_{\rm coh}$ is defined with a square root. \cite{ALICE:2023jgu,CMS:2023snh}

\begin{figure}
    \centering
    \includegraphics[scale=0.65]{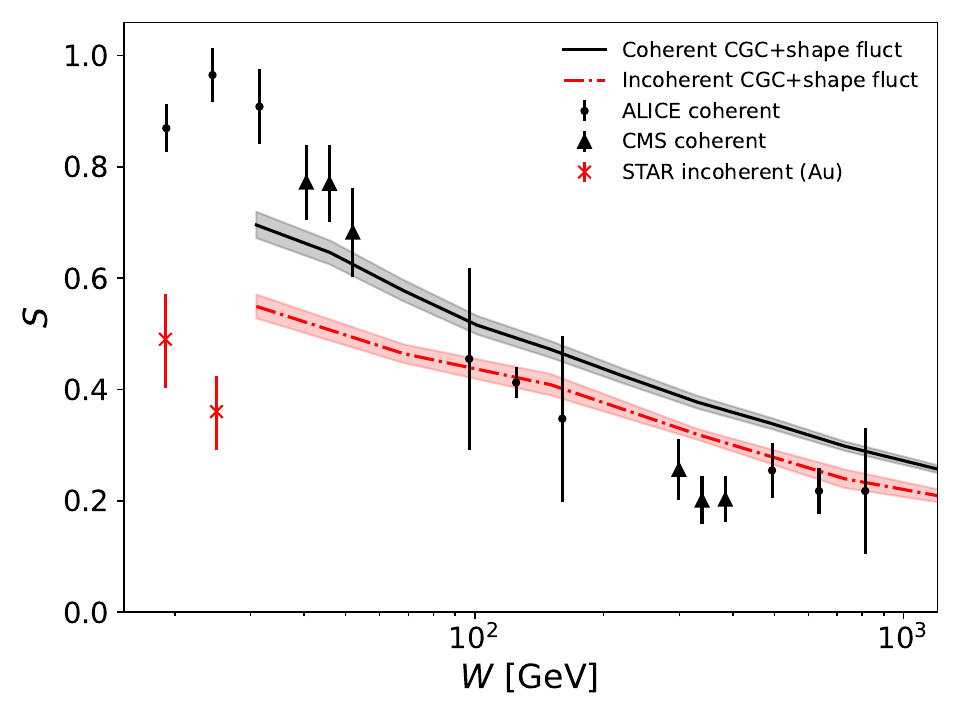}
    \caption{Suppression factors for coherent and incoherent production for Pb targets computed in the CGC framework with nucleon shape fluctuations \protect\cite{Mantysaari:2023xcu} and compared to the ALICE \protect\cite{ALICE:2023jgu}, CMS \protect\cite{CMS:2023snh}, and STAR \protect\cite{STAR:2023gpk} \protect\cite{STAR:2023nos} data (STAR data for Au targets). \label{fig:S}  } 
\end{figure}

For the incoherent cross section, one can define the suppression factor: \cite{STAR:2023gpk,STAR:2023nos}
\begin{equation}
    S_\mathrm{incoh} = \frac{ \sigma^{\gamma+ A \to V + A^*}}{A (\sigma^{\gamma + p \to V + p^*} + \sigma^{\gamma + p \to V + p} ) }\,.
    \label{eq:sincoh}
\end{equation}
We present results for the energy dependence of $S_{\rm coh}$ and $S_{\rm incoh}$ in Fig.\,\ref{fig:S}, comparing to results from ALICE and CMS for $S_{\rm coh}$ \cite{ALICE:2023jgu} \cite{CMS:2023snh} and from STAR for $S_{\rm incoh}$. \cite{STAR:2023gpk,STAR:2023nos}

The LHC data for the coherent suppression factor seems to have a steeper $W$ dependence than the model calculation. A slower evolution in heavy nuclei compared to the proton reference could improve agreement. One way to achieve this would be a $Q_s$ dependent running coupling in the evolution kernel. We note, however, that in our current setup the JIMWLK evolution includes the running coupling which depends on the daughter dipole sizes, and thus should indirectly depend on the saturation scale $Q_s$. We compare to STAR results for Au targets, where the kinematics leads to larger $x$ than the validity range of the framework. It will be interesting to compare to data for $S_\mathrm{incoh}$ at higher $W$ to test the prediction in Fig.\,\ref{fig:S}.

\section{Azimuthal Anisotropies from Interference}
In UPCs there is interference between the contributions where one nucleus is the photon source and the other is the target and the one where the roles are flipped. This can lead to interesting interference patterns. In fact, because the emitted photon is linearly polarized, the interference leads to characteristic azimuthal modulations in the angle between the decay products of a produced vector meson. This was measured for example in diffractive $\rho$ production and its decay into pions at STAR \cite{STAR:2022wfe}.

The cross section for this process can be computed in the CGC framework using a joint impact parameter and transverse
momentum-dependent formulation \cite{Xing:2020hwh,Hagiwara:2020juc}.
Here, the differential cross-section for the diffractive $\rho$ (or $\phi$) production, and the subsequent decay into pions or kaons,
respectively, is given by \cite{Mantysaari:2023prg}
\begin{align}
    \frac{\der \sigma^{ \rho \to \pi^+ \pi^- (\phi\to K^+ K^-)}}{\der^2 \Pt \der^2 \qt \der y_1 \der y_2 }  
     = \frac{1}{4 (2\pi)^3} \frac{P_\perp^2 f^2}{(Q^2 -M_V^2)^2 + M_V^2 \Gamma^2}  \Bigg \{ C_0(x_1,x_2,|\qt|)  +  C_2(x_1,x_2,|\qt|) \cos(2 (\phi_{\Pt} -\phi_{\qt})) \Bigg \} \,,
\end{align}
where $\Pt =(\textbf{p}_{1\perp}-\textbf{p}_{2\perp})/2$ and $\qt = \textbf{p}_{1\perp} + \textbf{p}_{2\perp}$ with $\textbf{p}_{1\perp}$ and $\textbf{p}_{2\perp}$ being the transverse momenta of the measured decay particles. Further, $Q$ is the invariant mass of the daughter particle system, and $y_1$ and $y_2$ are the daughter particles’ rapidities. The decay width is $\Gamma$ and $f$ are effective couplings.
Here, we have already separated the cross section into an isotropic piece and a $\cos(2\phi)$ modulation. The coefficients of the two components are
\begin{align}
 C_0(x_1,x_2,|\qt|) &=  \left<\int \der^2 \Bt \Mcal^i(x_1,x_2,\qt,\Bt) \Mcal^{\dagger,i} (x_1,x_2,\qt,\Bt) \Theta(|\Bt| - B_{\rm min})\right>_{\Omega}  \label{eq:C0} \,,\ \text{and} \\
 C_2(x_1,x_2,|\qt|) &= \left( \frac{2 \qt^i \qt^j}{\qt^2} - \delta^{ij} \right) \left<\int \der^2 \Bt \Mcal^i(x_1,x_2,\qt,\Bt) \Mcal^{\dagger,j} (x_1,x_2,\qt,\Bt) \Theta(|\Bt| - B_{\rm min}) \right>_{\Omega}\label{eq:C2} \,.
\end{align}
The amplitudes $\Mcal^{i}(x_1, x_2,\qt,\Bt)$ are expressed as the convolution of the photon field $ {\Fcal}_{A_k}^i(x_k,\bt)$ with the vector meson production amplitude $ {\Acal}_{A_k}(x_k,\bt)$ in coordinate space \cite{Mantysaari:2022sux}
\begin{align}
     \Mcal^i(x_1,x_2,\qt,\Bt) 
     = \int \der^2 \bt e^{-i\qt\cdot\bt} \left[  {\Acal}_{A_1}(x_1,\bt)  {\Fcal}_{A_2}^i(x_2,\bt-\Bt) + 
      {\Acal}_{A_2}(x_2,\bt-\Bt)  {\Fcal}_{A_1}^i(x_1,\bt) \right].
     \label{eq:amplitude_coordinate_space}
\end{align}
The vector meson production amplitude is computed in the CGC framework using Eqs.\,\eqref{eq:Amp} and \eqref{eq:CGCdip}, and the photon field is given by 
\begin{align}
    \Fcal_A^j(x,\Bt)  =\frac{1}{2\pi} \frac{Z \alpha_\mathrm{em}^{1/2} \omega}{\pi\gamma} \frac{\Bt^j}{|\Bt|} K_1\left( \frac{\omega|\Bt|}{\gamma}\right) \,.
    \label{eq:EM_fieldZ_Gauss}
\end{align}
Note that we recover the photon flux in Eq.\,\eqref{eq:flux} from 
\begin{align}
    n(\omega,|\Bt|)= (2\pi)^2 \left|\Fcal_A(x,|\Bt|)\right|^2\,.
\end{align}

\begin{figure}
    \centering
    \includegraphics[scale=0.39]{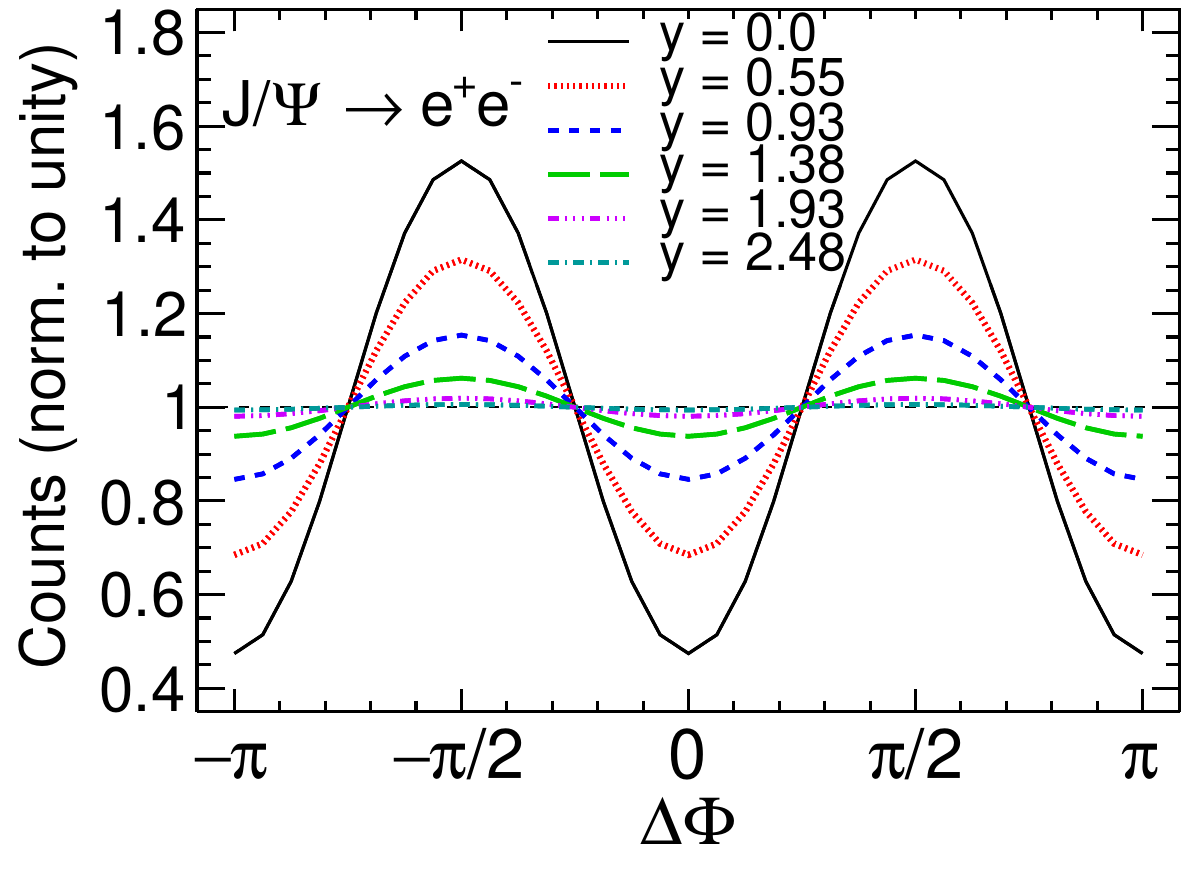} \hfill
    \includegraphics[scale=0.39]{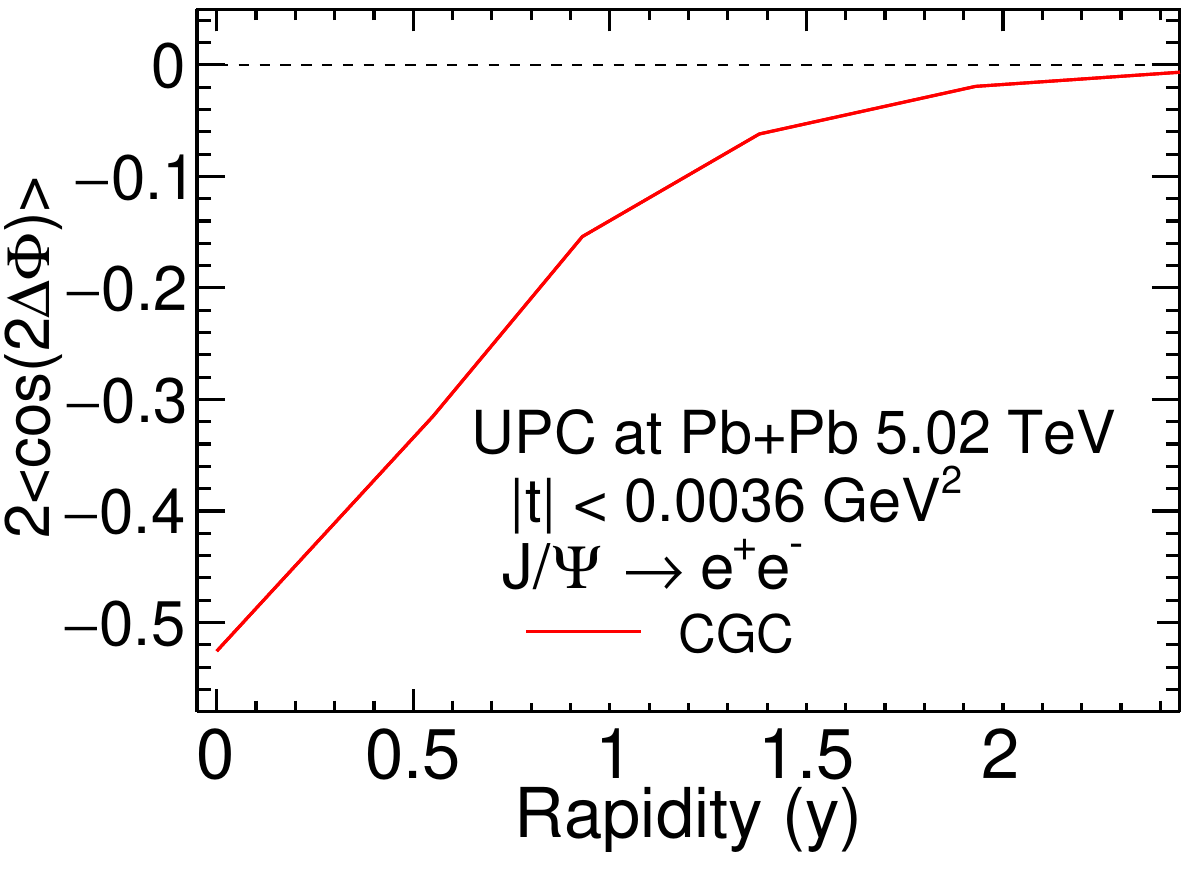}
    \caption{Left: The modulation in the azimuthal angle difference $\Delta \Phi =\phi_{\Pt} -\phi_{\qt}$ in $J/\psi$ production in UPCs at the LHC for different $J/\psi$ rapidities $y$. Right: The elliptic anisotropy coefficient as a function of rapidity extracted from the left plot.  
    \label{fig:mod}  } 
\end{figure}

For $J/\psi$ production and its decay in to dileptons, the cross section takes the form \cite{Mantysaari:2023prg,Brandenburg:2022jgr}
\begin{align}
    &\frac{\der \sigma^{ \jpsim \to l^+ l^- }}{\der^2 \Pt \der^2 \qt \der y_1 \der y_2 }  = \nonumber \\ 
    & ~ \frac{24 \alpha_{\mathrm{em}}^2 e_q^2}{(Q^2 -M_V^2)^2 + M_V^2 \Gamma^2} \frac{|\phi_{\jpsim}(0)|^2}{\pi M_V}  \Bigg \{ \left[ 1 - \frac{2 \Pt^2}{M_V^2} \right] C_0(x_1,x_2,|\qt|)  - \frac{2 \Pt^2}{M_V^2} C_2(x_1,x_2,|\qt|) \cos(2 (\phi_{\Pt} -\phi_{\qt})) \Bigg \}\,,
    \label{eq:Jpsi-cross-section}
\end{align}
where $|\phi_{\jpsim}(0)|^2 = 0.0447~{\rm GeV^3}$ \cite{ParticleDataGroup:2018ovx} is the value of the modulus squared of the radial wave function of the \jpsi at the origin.  The \jpsi EM decay width into two leptons is related to this via $\Gamma=16 \pi \aem^2 e_q^2 \frac{|\phi_{\jpsim}(0)|^2}{M_V^2} $ \cite{ParticleDataGroup:2018ovx}, with $e_q= 2/3$ the charm quark charge in units of $e$.

Many results for azimuthal anisotropies computed in the CGC framework and compared to data from the STAR Collaboration have been published. \cite{Xing:2020hwh,Hagiwara:2020juc,Brandenburg:2022jgr,Mantysaari:2023prg} Here we concentrate on the rapidity dependence of the interference effect, presented in Fig.\,\ref{fig:mod}. The figure shows results for the azimuthal anisotropy in $\phi_{\Pt} -\phi_{\qt}$ of the leptonic decay products in $J/\psi$ production in $\sqrt{s_{NN}}=5.02\,{\rm TeV}$ Pb+Pb UPCs. Plotting counts as a funciton of $\Delta\Phi= \phi_{\Pt} -\phi_{\qt}$ for different $J/\psi$ rapidities $y$, one can see that with increasing rapidity the modulation becomes weaker, because the interference decreases as the two interfering processes become increasingly distinguishable. The right panel shows the extracted elliptic anisotropy coefficient $2\langle \cos(2\Delta \Phi)\rangle$ as a function of rapidity. We note that especially in the case of light decay products such as $e^+$ and $e^-$, corrections from soft photon radiation are expected. They were shown to predominantly affect results at $q_{\perp} > 0.12~{\rm GeV/c}$ \cite{Brandenburg:2022jgr}.

\section{Nuclear structure effects}
As diffractive vector meson production is sensitive to the geometry
of the target, and via incoherent production also its fluctuations, important information on nuclear deformation, clustering and subnucleonic structure can be extracted by measuring the $t$-differential cross sections in $\gamma+A$ collisions. 

For example, the deformation of a nucleus can be incorporated via the Woods-Saxon parametrization
\begin{equation}\label{eq:WS}
    \rho(r,\theta) = \frac{\rho}{1+\exp[(r-R'(\theta))/a_{\rm WS}]}\,,
\end{equation}
with $R'(\theta)=R_{\rm WS}[1+\beta_2 Y_2^0(\theta)+\beta_3 Y_3^0(\theta) +\beta_4 Y_4^0(\theta)]$, and $\rho$ is the nuclear density at the center of the nucleus. Here $R_{\rm WS}$ is the radius parameter, $a_{\rm WS}$ is the skin diffuseness, and $\theta$ is the polar angle. The spherical harmonic functions $Y_l^m(\theta)$ and the parameters $\beta_i$ account for possible deformations.

\begin{figure}
    \centering
    \includegraphics[scale=0.5]{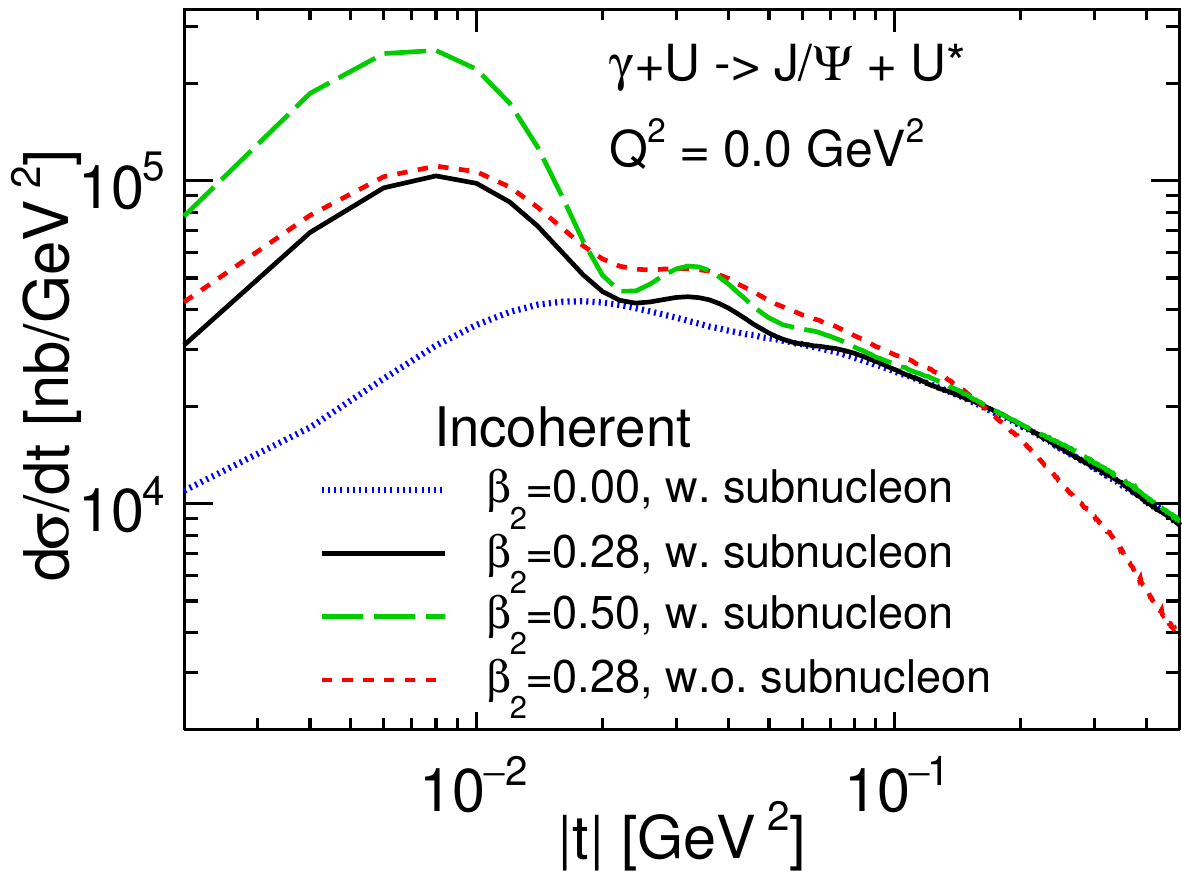}
    \caption{Differential incoherent $J/\psi$ photoproduction cross sections at $x=1.7 \times 10^{-3}$ for different $\beta_2$ values and with or without subnucleonic fluctuations. \label{fig:eU}  } 
\end{figure}

Some effects of deformation and subnucleonic structure are summarized in Fig.\,\ref{fig:eU}, where the incoherent cross section for $J/\psi$ production at $Q^2=0$ in $\gamma+U$ collisions is shown. Including subnucleon fluctuations, we show the result for the default quadrupole deformation $\beta_2=0.28$ along with results for a fictional uranium nucleus with $\beta_2 = 0$ and $\beta_2=0.5$. We also present the result for the default uranium target without subnucleonic fluctuations. The effects of varying $\beta_2$ is significant at $|t| < 2\times 10^{-2}\,{\rm GeV^2}$, while subnucleon fluctuations affect the cross section at $|t| > 0.2 \,{\rm GeV^2}$. It has been shown that varying $\beta_i$ for increasing $i$ affects the cross section at increasing $|t|$, showing that such measurement has sensitivity to nuclear structure over a large range of length scales. \cite{Mantysaari:2023qsq} This could be explored at a future Electron Ion Collider (EIC), but also in UPCs with a variety of nuclei. 

Nuclear structure also affects the azimuthal modulations driven by interference that were discussed in the previous section. Here, it was shown that in particular the modification of $B_{\rm min}$ by nuclear radius and deformation affects the amplitude of the modulation \cite{Mantysaari:2023prg}.

\section{Conclusions and Outlook}
Vector meson production in ultraperipheral heavy ion collisions is an excellent tool to study effects of nuclear shadowing and saturation. It further provides important information on nuclear geometry, including radii, deformations, and substructure. We compared many different models that have been employed to describe experimental data on coherent vector meson production in UPCs, and laid out some similarities and differences. We delved into some details of the Color Glass Condensate calculations and highlighted the energy dependence of nuclear suppression from QCD evolution, azimuthal anisotropies from interference effects in UPCs, and the effects of nuclear structure.

All results presented here are leading order (LO) calculations. The field is moving towards higher precision with next-to-leading order (NLO) calculations, both in the perturbative QCD (pQCD) collinear factorization framework \cite{Eskola:2022vpi}, and the dipole picture \cite{Escobedo:2019bxn,Mantysaari:2021ryb,Mantysaari:2022kdm}. In the collinear factorization approach, an interesting observation at NLO was that whike the LO and NLO gluon amplitudes dominate over the NLO quark contribution, they cancel to a large degree, increasing the importance of the NLO quark contribution.

With increasingly precise calculations and more experimental data from RHIC and LHC, UPCs will play a significant role in our endeavor to pin down gluon saturation and explore the regime of non-linear QCD. This is not least due to the fact that in UPCs we can reach very low $x$ values and study a variety of nuclear targets. Besides exclusive dijet and more inclusive observables, diffractive vector meson production is an excellent tool to address exciting physics questions already now, before the EIC will go online.

\section*{Acknowledgments}
We thank Vadim Guzey for helpful comments on the manuscript.
We thank Dagmar Bendov\'a, Michal Broz, Jan \v{C}epila, Vadim Guzey, Spencer Klein, Agnieszka Luszczak, and Wolfgang Schafer for providing data and help with generating Figure \ref{fig:ALICEplot}. 
This material is based upon work supported by the U.S. Department of Energy, Office of Science, Office of Nuclear Physics, under DOE Contract No.~DE-SC0012704 (B.P.S.) and Award No.~DE-SC0021969 (C.S.), and within the framework of the Saturated Glue (SURGE) Topical Theory Collaboration.
C.S. acknowledges a DOE Office of Science Early Career Award. 
H.M. is supported by the Research Council of Finland, the Centre of Excellence in Quark Matter, and projects 338263 and 346567, and under the European Union’s Horizon 2020 research and innovation programme by the European Research Council (ERC, grant agreement No. ERC-2018-ADG-835105 YoctoLHC) and by the STRONG-2020 project (grant agreement No 824093).
F.S. is supported by the U.S. DOE Office of Nuclear Physics under Grant No. DE-FG02-00ER41132.
W.B.Z. is supported by DOE under Contract No. DE-AC02-05CH11231, by NSF under Grant No. OAC-2004571 within the X-SCAPE Collaboration, and within the framework of the SURGE Topical Theory Collaboration. The content of this article does not reflect the official opinion of the European Union and responsibility for the information and views expressed therein lies entirely with the authors.
This research was done using resources provided by the Open Science Grid (OSG)~\cite{Pordes:2007zzb,Sfiligoi:2009cct}, which is supported by the National Science Foundation award \#2030508.








\section*{References}

\bibliography{refs}

\end{document}